\documentclass[twocolumn,final]{elsart5p}

\usepackage{graphicx}

\usepackage{amssymb}
\usepackage{array}
\begin{document}

\begin{frontmatter}


\title{Impact-driven effects in crystal growth: steering and transient mobility
at the Ag(110) surface}
\author{M. Ceriotti}
\address{Dipartimento di Scienza dei Materiali della Universit\`a degli Studi di Milano-Bicocca,
Via Cozzi 53, I-20125 Milano, Italy}
\author{R. Ferrando\corauthref{cor}}
\address{Dipartimento di Fisica della Universit\`a degli Studi di Genova,
Via Dodecaneso 33, I-16146 Genova, Italy}
and
\author{F. Montalenti}
\address{CNR/CNISM and L-NESS, Dipartimento di Scienza dei Materiali della
Universit\`a degli Studi di Milano-Bicocca, Via Cozzi 53, I-20125
Milano, Italy}
\corauth[cor]{Corresponding author. E-mail: ferrando@fisica.unige.it}

\begin{abstract}
Low-energy atomic impacts on the Ag(110) surface are investigated by molecular dynamics simulations based on reliableù
many-body semiempirical potentials.
Trajectory deflections (steering) caused by the atom-surface interaction
are observed, together with impact-following, transient-mobility effects.
Such processes are quantitatively analyzed and their dependence on the initial kinetic
energy and on the impinging direction is discussed. A clear influence of
the surface anisotropy on both steering and transient mobility effects is
revealed by our simulations for the simple isolated-atom case and in the submonolayer-growth regime. 
For the latter case, we illustrate the dependence of the distribution of adatoms, addimers, and larger islands 
on the deposition conditions.
\end{abstract}

\end{frontmatter}

\section{Introduction}
\label{sec::intro}

In 1989 Egelhoff and Jacob \cite{egelhoff89} reported a surprising result:
clear RHEED oscillations were in fact detected by the authors while monitoring metal
on metal epitaxy on fcc(001) surfaces, at temperatures as low as 77K. This evidence was unexpected since
at temperatures at which adatom diffusion is supposed to be totally frozen one would not
have predicted to witness ordered, layer by layer like growth. A possibile explanation was
offered by the authors: the condensation energy gained by bringing the impinging atom
on the surface was somehow transformed into kinetic energy of the adatom, allowing for
fast surface diffusion even when the substrate temperature was way too low to overtake any of the
relevant barriers. The intriguing interpretation triggered several theoretical and
computational works aimed at shedding light on the phenomenon. Molecular dynamics simulations
demonstrated that the lateral transient mobility acquired by the impinging atoms is usually limited
to a few atomic spacings, the condensation energy being fast distributed among the outermost
substrate atoms \cite{evans90,sanders92}. Nonetheless, a different kind of transient mobility
was shown to play a key role in determining a certain degree of ordering in growth. Newly deposited atoms
hitting  small three-dimensional features (e.g. pyramids) eventually present on the surface
were shown to easily (the process being barriereless in some cases) funnel down \cite{evans90,evans91}
at the foot of the protrusion until a stable site with the correct coordination was found. The effect
was demonstrated to produce a significant smoothening of the growing film \cite{evans91}.

More recently a different kind of deposition-induced mechanism was observed, leading, instead,
to enhanced roughening of the growing surface. A set of nice experimental observations on copper
growth on Cu(001) at low incidence angles \cite{vandijken99,vandijken00} showed that the long-range
interaction between the incoming atoms and the surface can lead to a significant raise in the rougness of the
film and in the slope of the typical mounds. Atomistic  simulations \cite{montalentiprb01,montalentiprl01}
confirmed this steering
effect, and showed that even under usual normal-incidence conditions, a 3D protrusion eventually formed
during growth tends to deflect incoming atoms towards it, thus enhancing its volume. Since a strong trajectory
deflection is possibile only for incoming atoms with a small velocity component in the direction
perpendicular to the surface,
steering is stronger when the initial kinetic energy and/or the incidence angle is smaller.
Steering was analyzed, in a slightly different context, also by Raible {\it et al.}, nicely
emphasizing the role played by this effect in inducing an instability
during growth \cite{raible00,raible02}. Incoming-atom trajectory deflection received further attention in the
last few years, when a detailed theoretical analysis of the influence of both long-range
and short-range atom-surface interaction on islands shape and slope was presented
\cite{amar02,amar03,amar04,amarbogicevic04},
showing that the picture of funneling-dominated low-temperature growth, due to steering, is not sufficient to
capture the whole complexity of the physics involved.

In the present work we analyze transient mobility and steering effects in Ag/Ag(110). Our aim is to
understand if the strong surface anisotropy can influence impact-following events, making it possible to tune
some of the characteristics of the growing film not only by changing the initial kinetic energy and the
impact angle of the impinging atoms, but also their azimutal angle. We recall that diffusion and growth at
(110) fcc metal surfaces, including Ag(110), has been extensively investigated from both the experimental
\cite{prlauag,linderoth97,horch99,rusponi981,rusponi982,boragno98,molle04,pedemonte04,demongeot03,pedemonte03} and the
theoretical point of view \cite{ferrando96,hontinfinde96,montalentiprb99,montalentiprl99,pjf,kurpick01,ferrandokmc,videcoq02,zhu04,hontinfindecpl,ndongmouo04}.
A detailed analysis  of the impact-driven effects under different deposition conditions, however, is
still lacking in the literature.

In the next Section we shall describe the methodology used for the present study, while in Section
\ref{sec::single} results concerning isolated-atom deposition are discussed. If the effect of both steering
and transient mobility are already well evident and easily identified for this simple case, it is in
Section \ref{sec::sub} that their influence on growth is more directly shown. Submonolayer-growth
simulations are indeed analyzed, together with the statistics of deposition-dependent distributions of
adatoms, addimers, and larger one-dimensional islands. Finally, a critical discussion of the simulations is 
offered in Section \ref{sec::critical}, while the summary of the results and our conclusions
are reported in Section \ref{sec::conclusions}.

\section{Details of the calculations}
\label{sec::method}

The whole set of results here presented was obtained by using classical molecular dynamics (MD) simulations.
The Ag-Ag interaction was described by the Rosato-Guillop\'e-Legrand (RGL) semiempirical potentials \cite{rgl1},
characterized by a many-body form derived in the framework of the second-moment approximation of the tight-binding 
scheme. Although the RGL potentials in the present parametrization (the complete set of parameters is reported, 
e.g. in Ref. \cite{montalentiprb99}) were not fitted on surface properties, they were shown to yield results in 
terms of surface diffusivities in nice agreement with experiments
and/or ab initio calculations \cite{montalentiprb99,ferrandokmc,alanissila} not only for Ag,
but also for Cu and Au. Surface reconstructions are also nicely predicted \cite{rgl2}.
The simulation cell used to represent the pristine Ag(110) surface in the calculations described in 
Section \ref{sec::single}, dealing with isolated-atom
deposition only, was composed of 768 atoms (16 layers, 48 atoms per layer), while for the submonolayer-growth
simulations of Section \ref{sec::sub} an initial slab of 4320 atoms (20 layers, 216
atoms per layer) was used. Periodic boundary conditions were applied in the surface plane.
In the MD simulations all of the atoms were allowed to move freely but the three
bottom layers that were kept frozen to bulk positions. Additional atoms were
deposited only at the upper slab termination.

\begin{figure}
\centerline{\includegraphics*[width=1.0\columnwidth]{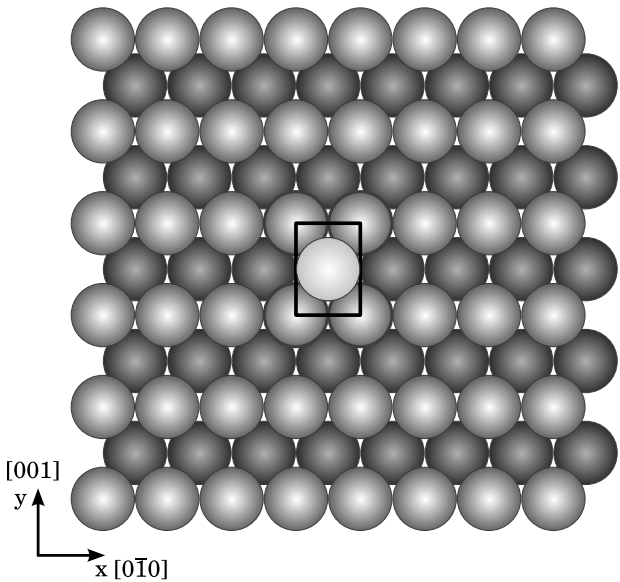}}
\caption{Top view of the simulation cell used in isolated-atom deposition simulations. A larger slab was used in submonolayer-growth simulations. The equilibrium position for a single adatom is also shown.}
\label{fig::slab}
\end{figure}

A top view of the smaller simulation cell is represented in Fig.\ref{fig::slab}. The presence of channels in the 
$\lbrack 0 \bar 1 0 \rbrack$ direction  is evident. As usual in the description of such surface, we shall call 
in-channel (or, simply, $x$) the $\lbrack 0 \bar 1 0\rbrack$ direction and cross-channel ($y$) the [001] one.
Deposition of additional atoms and system equilibration was simulated using the following procedure. 
The three layers directly above the three frozen ones were
coupled to a Langevin thermostat (we used a friction term $\eta=5\times 10^{12}s^{-1}$) imposing a desired temperature, 
while the dynamics of all the other atoms was simulated by using a
standard velocity Verlet algorithm. The time step was set to 5fs, a value that
guarantees satisfactory energy conservation (in the absence of the thermostat).
In the single-atom deposition simulations the slab was first equilibrated for 3ps.
A new atom with the desired impact parameters (kinetic energy, normal and azimutal impact angles) was then introduced 
high above (right outside the cutoff radius) the surface. Due to the low impact-energy values here considered, 
5ps of evolution turned out to
be sufficient to bring the atom to the surface, to have its excess energy dissipated into the simulation cell, 
and to equilibrate the whole system.
In the submonolayer-growth simulation, after the first atom was deposited, additional ones were added one by one, 
with a rate of about $10^9$ layers/s (1 atom
every 10 ps). Several independent simulations were carried out in order to collect
a significant statistics. For post-processing and analysis purposes, right before inserting a new atom, 
configurations were saved and optimized using a standard steepest-descent algorithm. 
Further details on the simulations, and in particular on initial atom positioning, are given
in the two following dedicated Sections.

\begin{figure}
\centerline{\includegraphics*[width=1.0\columnwidth]{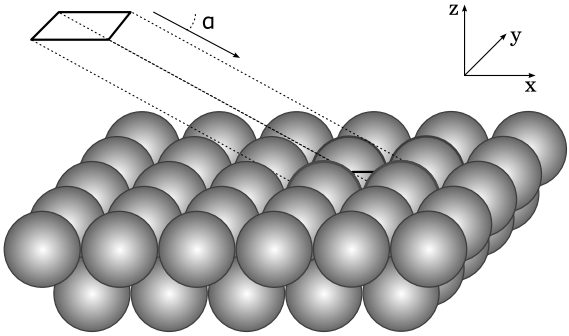}}
\caption{Schematic representation of the geometry for single atom deposition. The trajectories start 
from a randomly chosen location within the projection of the ``target'' cell high above the surface,
which serves as a reference point to compare the results for different geometries.}
\label{fig::storta}
\end{figure}

\section{Results: isolated-atom deposition}
\label{sec::single}

\begin{figure}
\parbox[t]{0.4\columnwidth}{\flushleft\includegraphics*[width=0.43\columnwidth]{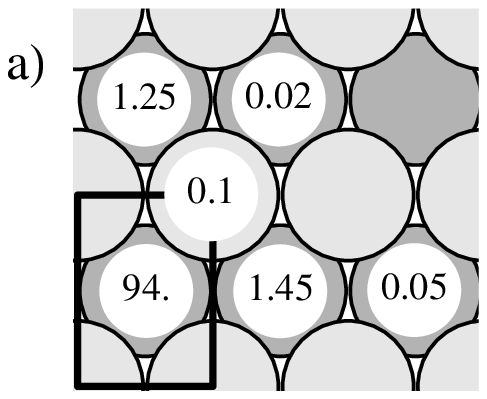}}
\parbox[t]{0.6\columnwidth}{\flushright\includegraphics*[width=0.55\columnwidth]{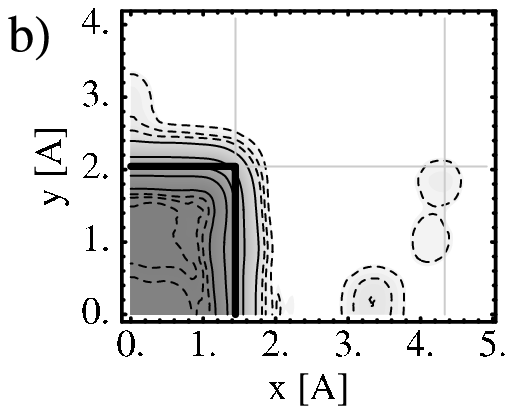}}

\parbox[t]{0.4\columnwidth}{\flushleft\includegraphics*[width=0.43\columnwidth]{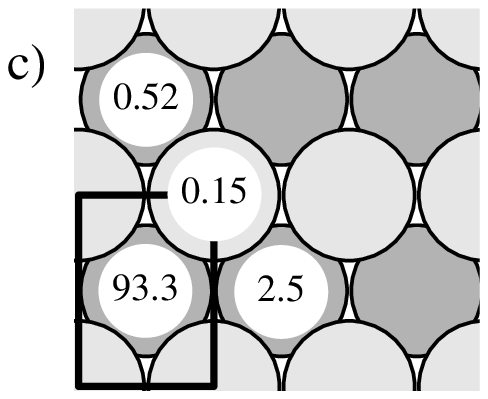}}
\parbox[t]{0.6\columnwidth}{\flushright\includegraphics*[width=0.55\columnwidth]{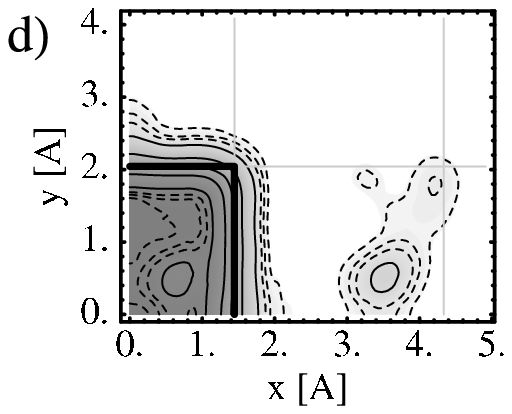}}

\caption{Figures show the final position of the impinging atom [a) and c)] and the 
capture probability function [b) and d)] for normal incidence 
simulations; impact energies are $0.1$eV [a) and b)] 
and at $1.0$eV [c) and d)].
The percentage of the trajectories ending up at each site is indicated, 
and $p(x,y)$ is drawn as a contourplot, where the continuous lines correspond 
to $0.1$, $0.3$, $0.5$, $0.7$, $0.9$, and the 
dashed lines to $0.01$, $0.03$, $0.05$, $0.95$, $0.97$, $0.99$
isoprobability contours. 
Only the relevant symmetry independent part of the simulation box is shown,
and the target unit cell is highlighted with a thick outline.
 }
\label{fig::atom-z}
\end{figure}

In order to investigate steering and transient mobility effects we started by analyzing the simplest case, 
i.e. the deposition of a single atom on the clean
Ag(110) surface. For the normal-incidence case, the overall effect of trajectory deflections can be monitored 
by simply running several independent simulations where the initial 
impinging-atom $(x,y)$ coordinates are randomly chosen within the surface unit cell, while the 
height $z$ above the surface is kept fixed at the cutoff radius value. The spread of the 
atom positions after the impact across and out of the unit cell are then analyzed and related 
to the deflections or to other atomic-scale mechanisms directly triggered by the impact. In the 
most general case where the impact is not normal and the azimutal angle is different than 
zero, the $(x,y)$ grid of starting coordinates, instead, was chosen over a displaced unit cell 
(the $z$ coordinate, again, corresponding to the cutoff radius), 
built along the initial direction of impact. The procedure is schematized in Fig.\ref{fig::storta}. 
This choice allowed us to directly compare the normal and out-of-normal 
deposition cases in terms of the final atomic positions, keeping the same  
surface unit cell as reference. Let us now analyze the results, which were obtained for a substrate
temperature T=100K. We shall comment further on the meaning of the subtrate temperature in our MD 
simulations in Section \ref{sec::critical}.
We performed simulations at normal and grazing ($\alpha = 12.5°$ with respect to Fig.\ref{fig::storta}) 
incidence, both along the in-channel and cross-channel directions, with an initial kinetic energy of
 $0.1$ and $1.0$ eV. In each case we analized the final, optimized position of the impinging atom,
gathering statistical data on the frequency of the various possible configurations; moreover, 
exploiting the translational simmetry of the clean surface (if a trajectory starting at $(x,y)$ 
ends up on the right-hand neighbour of the target cell, it would have on average ended up
on the target cell if it had started at $(x-a,y)$, where $a$ is the lattice parameter) one can 
calculate the capture area of a cell, i.e. the function $p(x,y)$ returning the probability that
a trajectory ``geometrically aimed'' at the position $(x,y)$ with respect to the center of a given cell
will land up at the equilibrium position over that cell.
One could extimate $p(x,y)$ dividing the surface with a dense mesh, and counting how many of the 
trajectories pointed to a given interval $(x_i,x_{i+1})\times(y_j,y_j+1)$ finish over the 
target cell; the number of trajectories one should accumulate to obtain smooth results over a fine grid
is enormous: we have instead calculated 
\[p(\mathbf{x})\approx
\frac{\sum_{i_{target}}\delta(\mathbf{x}-\mathbf{x}_i)}
{\sum_{i}\delta(\mathbf{x}-\mathbf{x}_i)}
\]
where the delta functions have been replaced by normalized gaussians
\[g_{\sigma}(\mathbf{x})=\frac{1}{\pi \sigma^2}e^{-\mathbf{x}^2/\sigma^2}\]
with the width $\sigma = 0.3$\AA, chosen in order to obtain a smooth $p(\mathbf{x})$. Although 
the resulting capture probability is heavily smeared and a quantitative analysis would require
more intensive sampling, the qualitative features of $p(\mathbf{x})$ are conserved, and highlight
geometrical effects responsible of captures processes.
The normal-incidence simulations (Fig. \ref{fig::atom-z}) show no surprises: most of the trajectories
finish with the impinging atom in the equilibrium position of the target cell. 
Trajectories aimed on the borders of the target cell may occasionally get captured on the nearest
neighbouring cell, and there is a small probability for exchange processes, with the impinging
atom taking the place of a surface atom, which becomes an adatom in the nearby row; as clearly shown in 
the contour plots in figure  \ref{fig::atom-z},  there is a well defined zone for which the 
falling atom has a non vanishing probability (reaching 10\% for the $1.0$eV case) to ``bounce''
into the nearby cell on the same row.
 For the exchange and rebound processes we may speak of transient mobility, because the impact energy
activates diffusion processes otherwise inaccessible in this temperature regime. There are no qualitative
differences between the $0.1$eV and the $1.0$eV cases, because most of the impact energy is due
to the attractive interaction between the incoming atom and the surface.
 
\begin{figure*}
\parbox[t]{0.46\textwidth}{\flushleft\includegraphics*[width=0.46\textwidth]{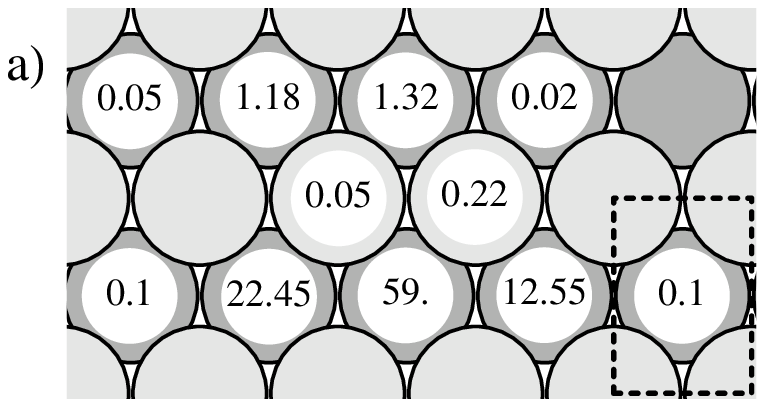}}
\parbox[t]{0.54\textwidth}{\flushright\includegraphics*[width=0.54\textwidth]{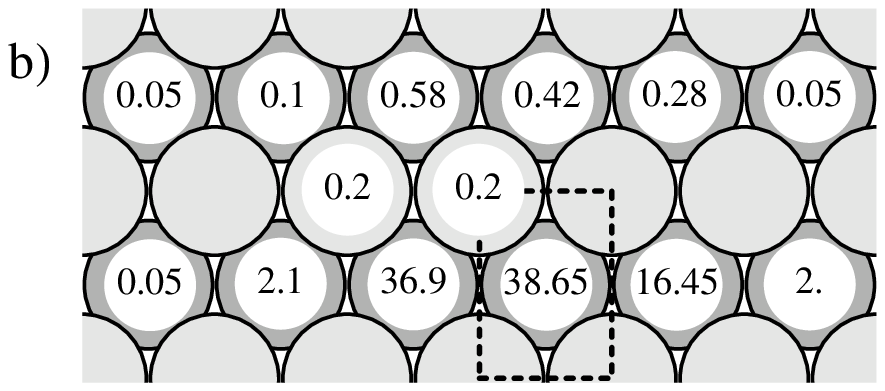}}
\\
\parbox[t]{0.46\textwidth}{\flushleft\includegraphics*[width=0.46\textwidth]{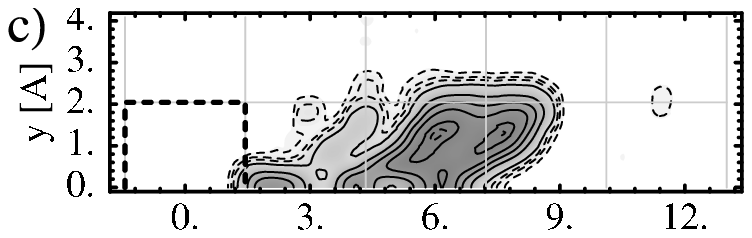}}
\parbox[t]{0.56\textwidth}{\flushright\includegraphics*[width=0.54\textwidth]{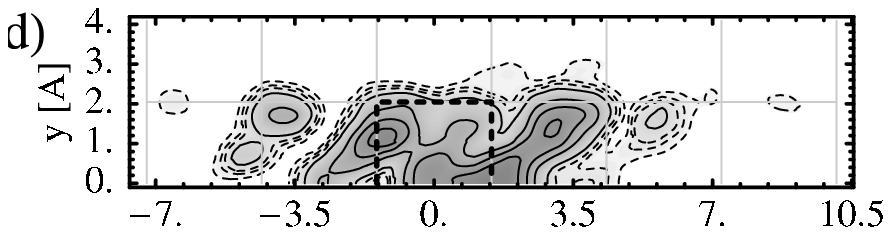}}

\caption{Figures show the final position of the impinging atom [a) and b)] and the capture 
probability function [c) and d)]  for in-channel grazing incidence ($\alpha=12.5°$, along $x$) simulations; 
impact energies are $0.1$eV [a) and c)] 
and at $1.0$eV [b) and d)].
The percentage of the trajectories ending up at each site is indicated, and $p(x,y)$ is drawn 
as a contourplot, where the continuous lines correspond to $0.1$, $0.3$, $0.5$, $0.7$, $0.9$, and the 
dashed lines to $0.01$, $0.03$, $0.05$, $0.95$, $0.97$, $0.99$
isoprobability contours. Only the relevant symmetry independent part of the simulation box is shown; the target unit cell
is four cells to the right [a) and b)] or to the left [c) and d)] of the reference cell, marked with 
a dashed outline.
 }
\label{fig::atom-x}
\end{figure*}

\begin{figure}
\centerline{\includegraphics*[width=0.8\columnwidth]{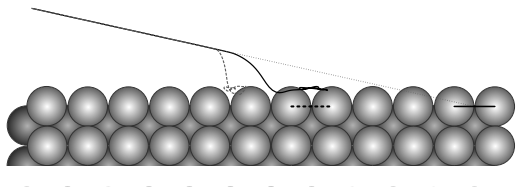}}
\caption{Two trajectories are shown, for grazing, in-channel incidence, with $0.1$ eV (dashed line) and 
$1.0$ eV (full line) initial kinetic energy. The dotted, straight line represent the path the incoming 
atom would follow in absence of any steering effects. The bold, full line outlines the target unit cell,
while the bold, dashed line outlines the reference cell used in figure~\ref{fig::atom-x}.}
\label{fig::steer-t}
\end{figure}

In figure \ref{fig::atom-x} the results for in-channel grazing incidence are shown; two effects
are to be considered to account for the results: on one hand, the initial velocity of the incident atom
has a big component parallel to the surface, which increases the in-channel mobility of the adatom 
immediately after the impact; on the other hand, steering causes the trajectory to digress from the
``geometrical'' one, causing the shooted atom to fall shorter than expected (Fig. \ref{fig::steer-t}). 
At such a low impact angle the steering effect is very pronounced, and for both initial energies the
atom settles down before it reaches the target cell; the atom travels longer distance at $1.0 eV$, not only because
the deflection from ideal trajectory is less pronounced, but also because of transient mobility effects,
as clearly depicted by the larger spread of capture probability. The impinging atoms can hop
a few times between in-channel adjacent sites,  depending
from the impact location and from the thermic displacements of the atoms it encounters along its trajectory.
The arrow-like shape of $p(x,y)$ displays clearly
that - on average - the falling atom travels half a lattice parameter farther if it lands into the ditch
than if it lands on the banks, both because it feels a bit later the steering effect due to interaction
with the surface, and because (thanks to the symmetric position into the groove) it dissipates slower
the momentum component along $x$.

\begin{figure*}
\parbox[t]{0.24\textwidth}{\flushleft\includegraphics*[width=0.24\textwidth]{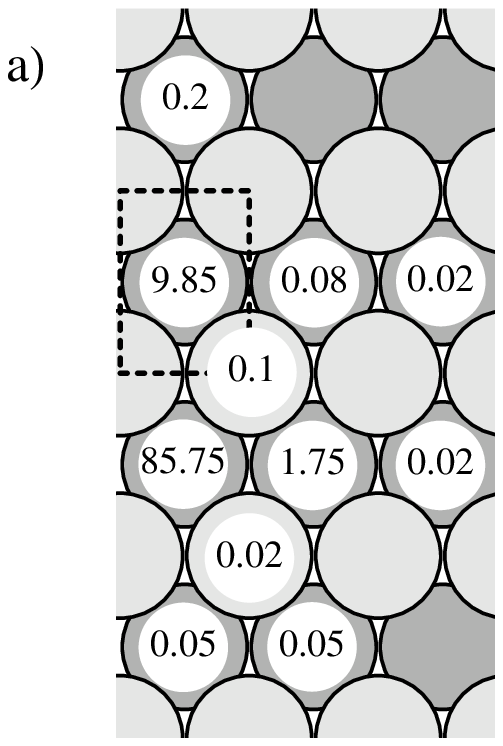}}
\parbox[t]{0.24\textwidth}{\flushleft\includegraphics*[width=0.24\textwidth]{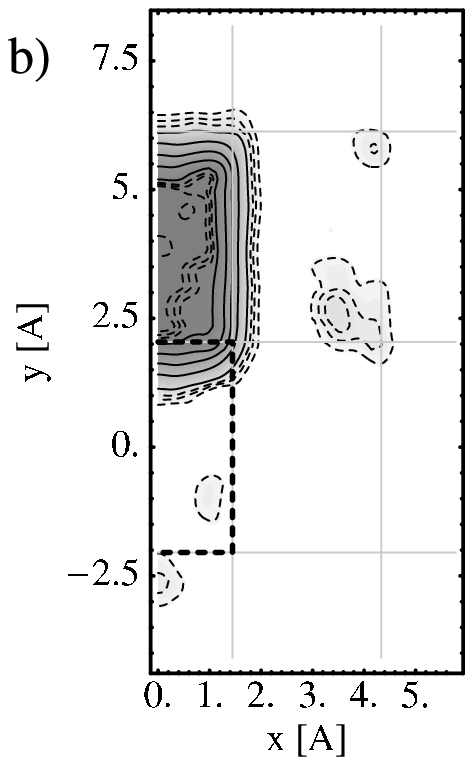}}
\parbox[t]{0.24\textwidth}{\flushright\includegraphics*[width=0.24\textwidth]{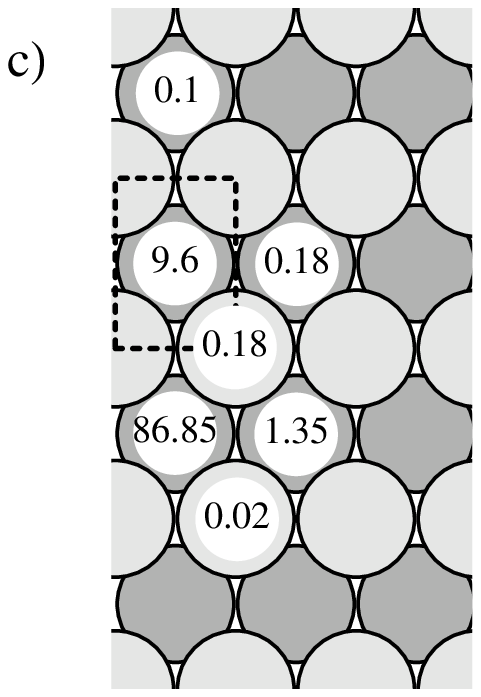}}
\parbox[t]{0.24\textwidth}{\flushright\includegraphics*[width=0.24\textwidth]{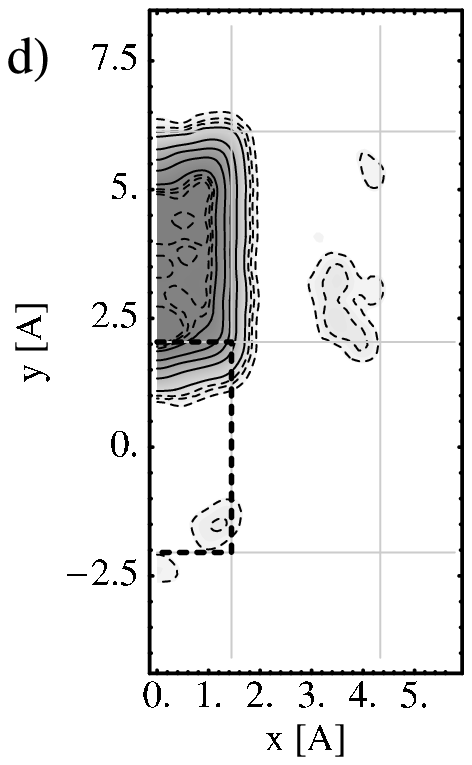}}

\caption{Figures show the final position of the impinging atom [a) and c)] and the 
capture probability function [b) and d)] for in-channel grazing incidence 
($\alpha=12.5°$, along $x$) simulations; impact energies are $0.1$eV [a) and b)] 
and at $1.0$eV [c) and d)].
The percentage of the trajectories ending up at each site is indicated, 
and $p(x,y)$ is drawn as a contourplot, where the continuous lines correspond 
to $0.1$, $0.3$, $0.5$, $0.7$, $0.9$, and the 
dashed lines to $0.01$, $0.03$, $0.05$, $0.95$, $0.97$, $0.99$
isoprobability contours. 
Only the relevant symmetry independent part of the simulation box is shown; the target unit cell
is three cells to the top [a) and c)] or to the bottom [b) and d)] of the reference cell, marked with 
a dashed outline.
}
\label{fig::atom-y}
\end{figure*}

The cross-channel grazing incidence results are depicted in figure \ref{fig::atom-y}; here we see again
steering effects that cause the impinging atom to fall sistematically short: since barriers for
cross-channel diffusion are much higher than those for in-channel diffusion, the transient mobility effects
are here much less evident than in the previous case, and the differences between the two initial 
kinetic energies are negligible, but for a slight enhancement of exchange events probability.

\section{Results: submonolayer growth}
\label{sec::sub}

If already the single-atom case revealed non-trivial impact-following events, it is
natural to predict that after one or more adatoms are already present on the
surface, impact-following events become more pronounced and complex at the
same time. Indeed, even a single adatom can be seen as a small protrusion extending out of the surface and, as such, 
steering trajectories of newly incoming material towards it. At the same time, the presence of an adatom within the 
channel can limit the range of transient-mobility diffusion. In order to look at similar effects, in
this Section we analyze simulations where, every $10$ ps, a new atom is deposited on the surface until a 
$0.5$ coverage is reached. To allow for some degrees of relaxation
(see Section \ref{sec::critical}), the substrate temperature was set to T=300K.
We performed simulations for the same six sets of impact parameters used in section \ref{sec::single},
but we selected for further analysis only the two in-channel grazing configurations and the $0.1$eV
normal-incidence one, because they better displayed the combined effects of steering and
transient mobility.  In order to better single out the effects of steering on growth, we compare the results
of our present MD simulations with those obtained by a Kinetic Monte Carlo (KMC) model \cite{videcoq02} at low temperature 
(100 K and below).
The KMC model incorporates all diffusive processes and random deposition with downward funnelling, but 
does not take into account the steering.

For each set of parameters we performed $30$ independent simulations up to $0.5$ coverage, and we
postprocessed the optimized structures after every impact to estimate
some averages, representative of the growing layer morphology, such as the average number of $n$-islands
at $\theta$ coverage, $a_n(\theta)$ or simply the average length of islands,
$l(\theta)=\sum_n{n a_n(\theta)}/\sum_n{a_n(\theta)}$. Statistical errors were estimated as the standard deviations
divided by the square root of the number of independent simulations.

\begin{figure}
\caption{(Color online) 
Average length of islands as a function of the coverage; results are averaged over 30
independent simulations, and dots represent error bars calculated as the standard deviation of
the data over $\sqrt{30}$. No error bars are shown for the model, as we were able to perform
over $10^3$ independent simulations, and error bars are negligible on the scale of the graphic.}
\vspace{0.2cm}\centerline{\includegraphics*[width=1.0\columnwidth]{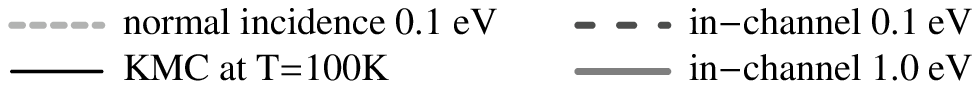}}\vspace{0.2cm}
\centerline{\includegraphics*[width=0.7\columnwidth]{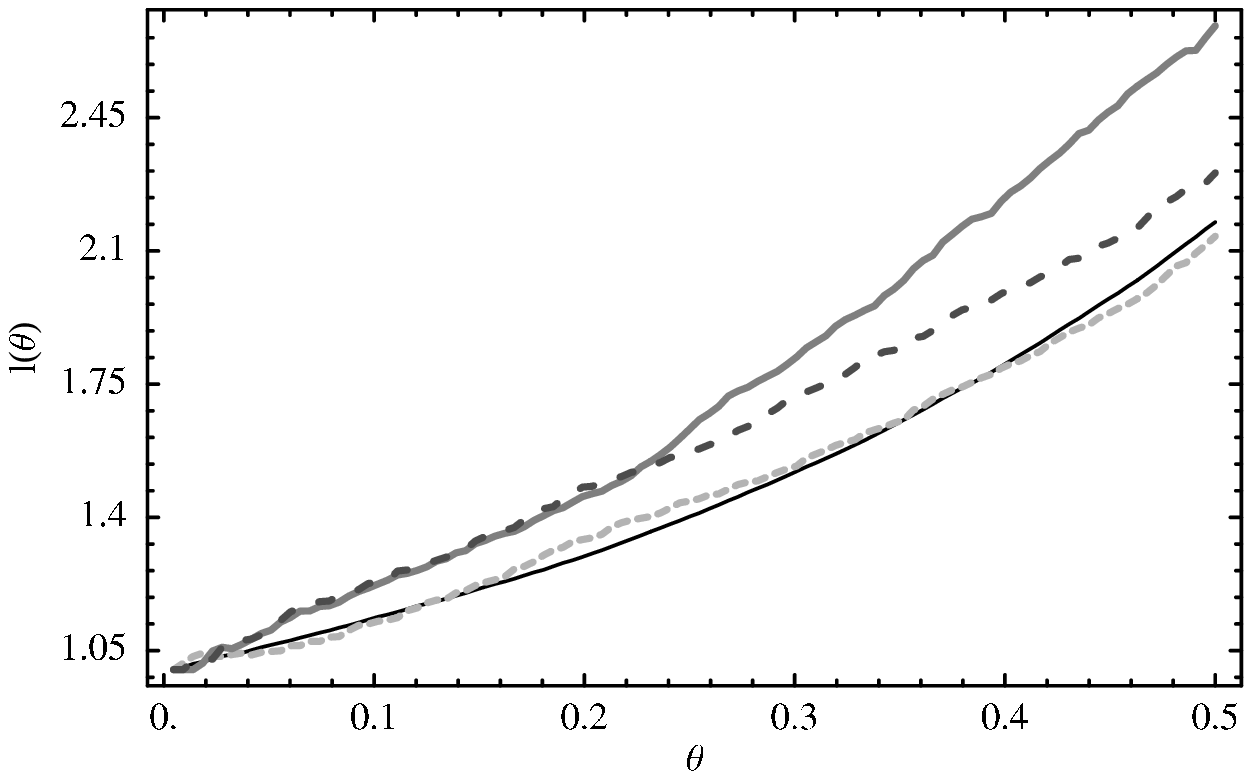}}
\label{fig::l-coverage}
\end{figure}

\begin{figure}
\caption{Average number of $n$-islands as a function of the coverage; results are averaged over 30
independent simulations, and dots represent error bars calculated as the standard deviation of
the data over $\sqrt{30}$. In the last panel, the number of atoms building up a second layer is 
drawn. 
}
\vspace{0.2cm}\centerline{\includegraphics*[width=1.0\columnwidth]{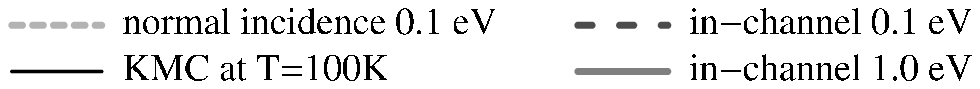}}\vspace{0.2cm}
\parbox[t]{0.49\columnwidth}{\flushleft\includegraphics*[width=0.49\columnwidth]{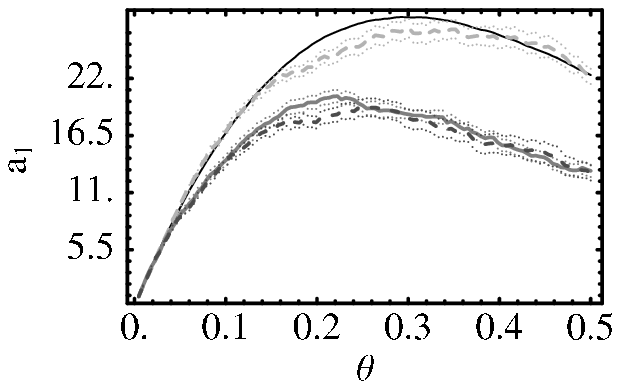}}
\parbox[t]{0.49\columnwidth}{\flushright\includegraphics*[width=0.49\columnwidth]{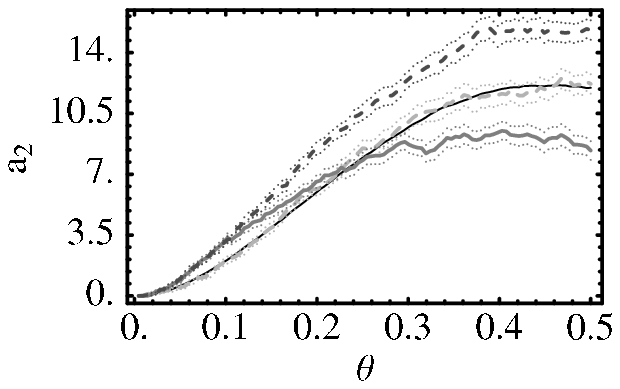}}

\parbox[t]{0.49\columnwidth}{\flushleft\includegraphics*[width=0.49\columnwidth]{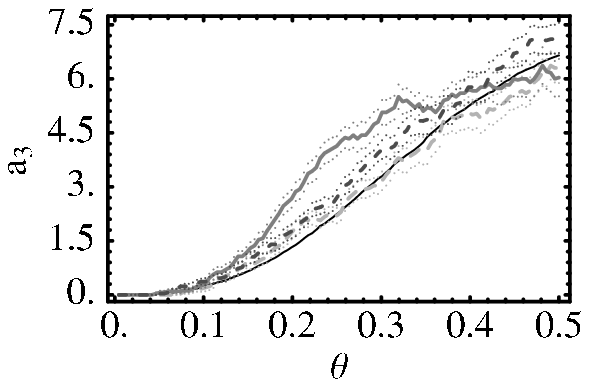}}
\parbox[t]{0.49\columnwidth}{\flushright\includegraphics*[width=0.49\columnwidth]{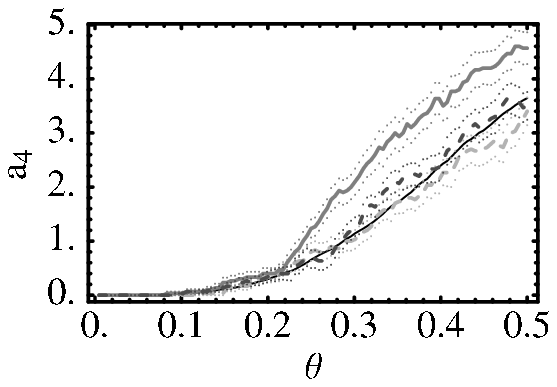}}

\parbox[t]{0.49\columnwidth}{\flushleft\includegraphics*[width=0.49\columnwidth]{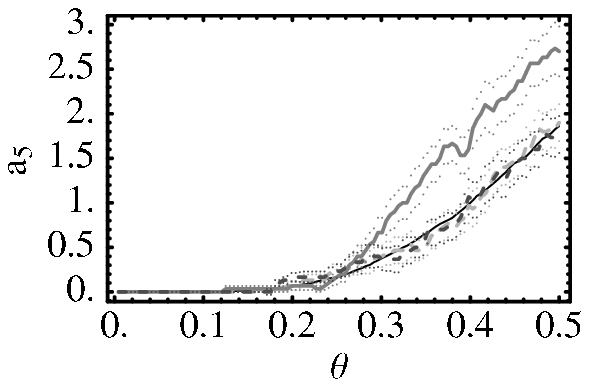}}
\parbox[t]{0.49\columnwidth}{\flushright\includegraphics*[width=0.49\columnwidth]{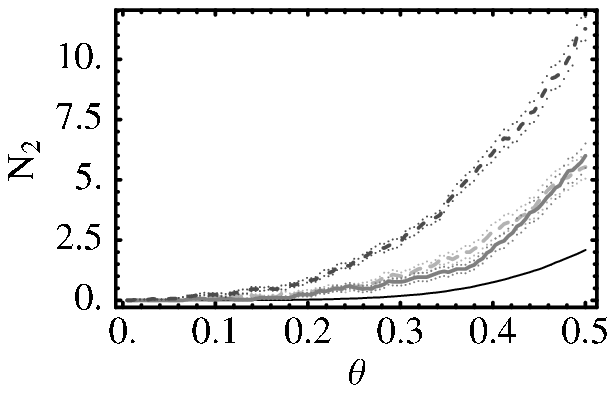}}
\label{fig::an}
\end{figure}

$l(\theta)$ is plotted in figure \ref{fig::l-coverage} and $a_n(\theta)$  from single adatoms up to $5$-islands
are reported in figure \ref{fig::an}; the film as grown under the different conditions has quite different
structures: the normal-incidence case compares very well with the KMC model, but for the
tendency to form 3-dimensional structures,  which is enhanced in the MD simulations. 
This shows that steering is effective in increasing the surface roughness at normal-incidence deposition.
The average island length is 30\% longer for 1.0eV in-channel grazing incidence, while the $0.1$eV
in-channel case is positioned halfway; in fact, $a_n(\theta)$ is a much more sensitive index of the film
structure, and displays for example numerous $\ge 4$ islands in the in-channel $1.0$eV configuration,
high abundance of dimers and tendency towards 3D growth in the $0.1$eV in-channel case.
Explaination for this variety in the in-channel trajectories requires again mutual influence
of transient mobility and steering: grazing incidence causes the impinging atom to interact for a
longer distance with the surface, feeling the effects of already deposited atoms, while momentum
parallel to the channels will determine if the impinging atom will be steered to fall right beside
another adatom (this explaining the high concentration of dimers in $0.1$eV case) or diffuse
into the channel until it knocks onto a growing island (this explaining the longer islands registered
at $1.0$eV.

\begin{figure*}
\caption{(Color online) Snapshots of the surface at $\theta = 0.5$ coverage, representative of the layer
morphology. Images represent some of the outcomes of our simulation, chosen for the best 
compliance with the average island size distribution. a) is from a normal incidence, $0.1$ eV simulation, b) and c) from grazing, in-channel simulations, respectively at $0.1$ and $1.0$ eV.}
\parbox[t]{0.32\textwidth}{\flushleft\includegraphics*[width=0.32\textwidth]{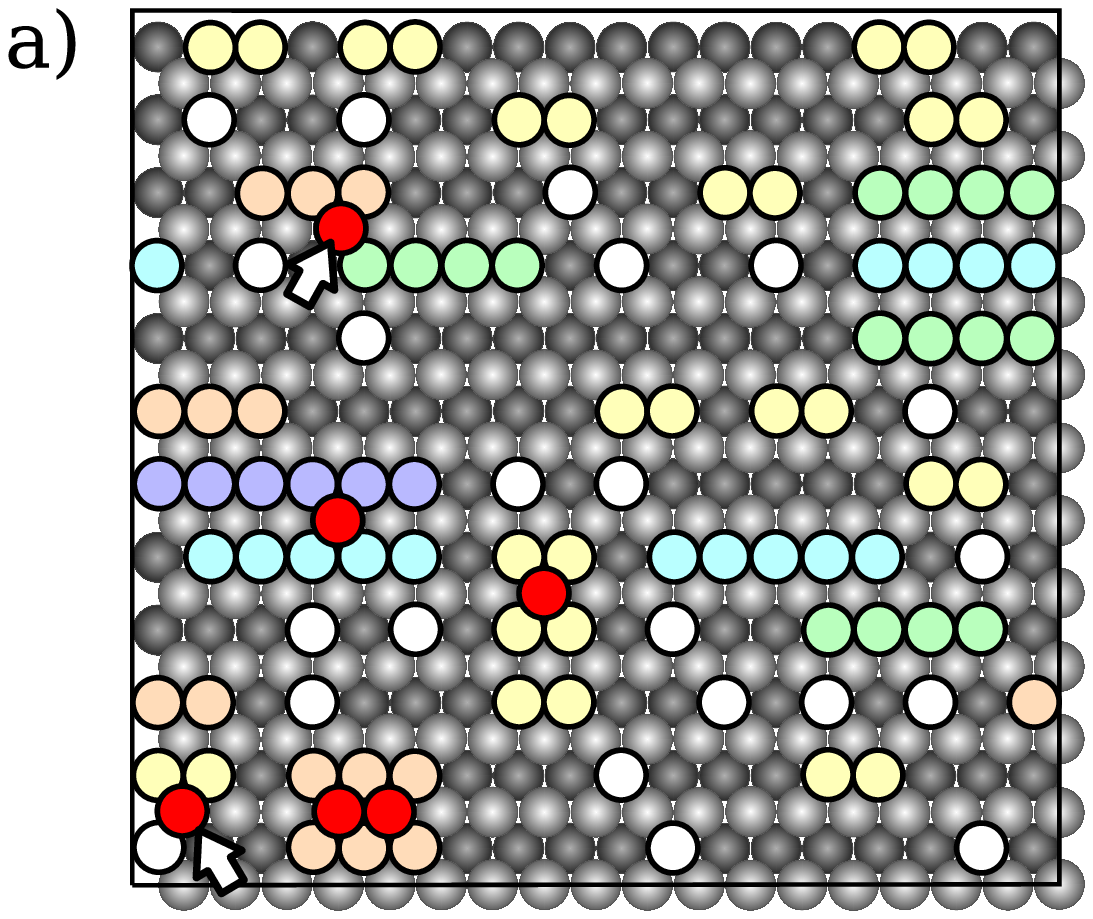}}
\parbox[t]{0.32\textwidth}{\flushleft\includegraphics*[width=0.32\textwidth]{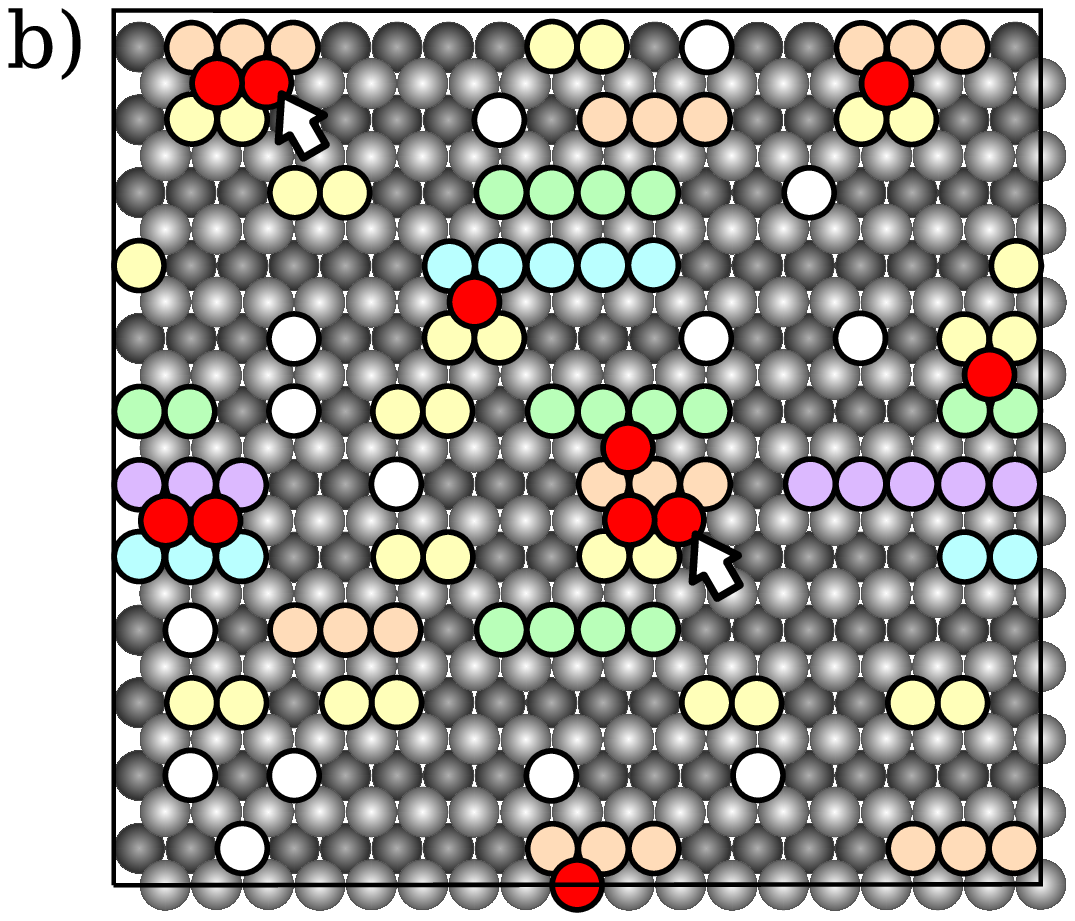}}
\parbox[t]{0.32\textwidth}{\flushright\includegraphics*[width=0.32\textwidth]{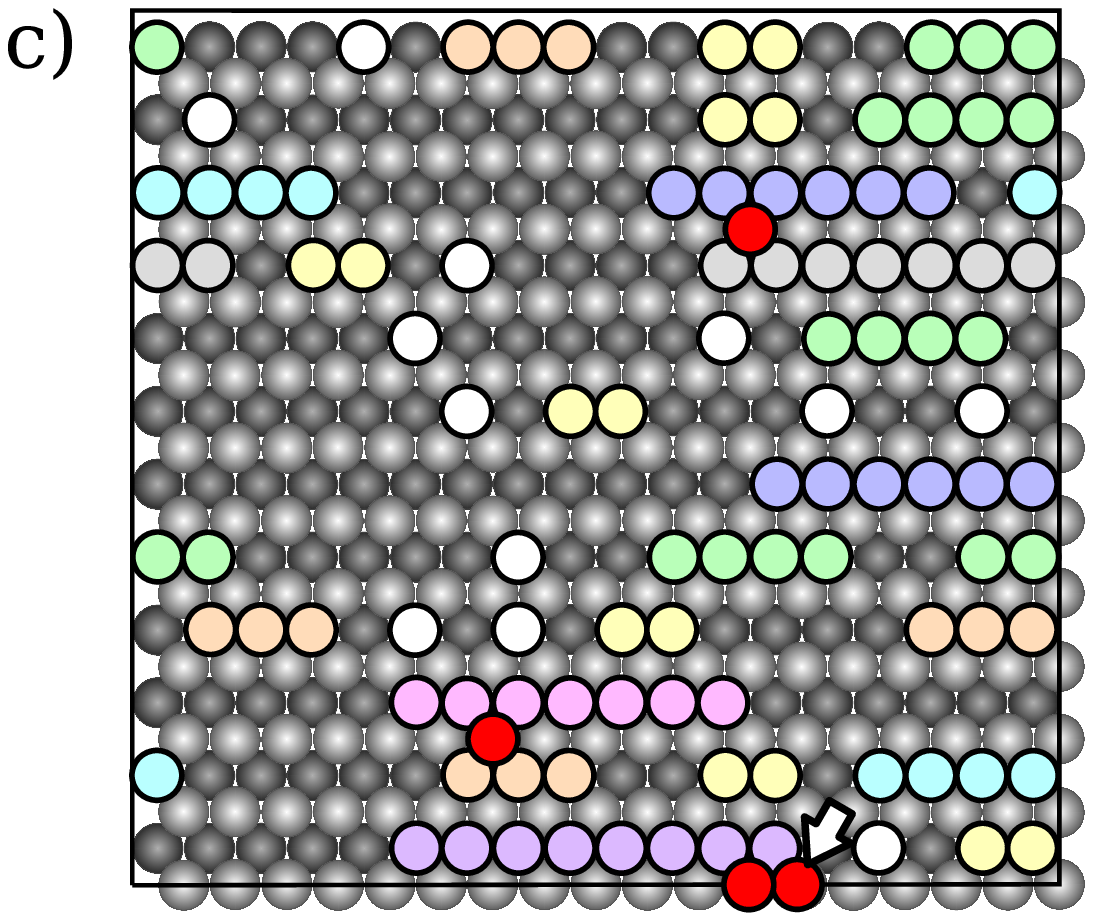}}
\label{fig::snap}
\end{figure*}

Our results show that surface roughness is smaller for high-energy grazing incidence than for
low-energy grazing incidence. This is due to the fact that, at  high impact energy, atoms landing on an already
formed island have sufficient transient mobility to reach the border of the island, and eventually surmount the low
in-channel Ehrlich-Schwoebel barrier \cite{videcoq02} to fall down on the lower terrace.

A careful calibration of the impact angle and energy would probably permit to fine-tune the
morphology, enhancing or depressing selectively the presence of the various types of islands.
The possibility to tune the 3-dimensional growth is also promising as it should have an
influence on the  defectivity of the film.

\section{Critical discussion: simulated substrate temperature}
\label{sec::critical}

MD simulations of Ag/Ag(110) growth in the submonolayer regime were carried out at a substrate temperature $T=300$K. 
This temperature value should not be regarded as representative of an actual experimental temperature. 
As explained here below, the present simulations are an attempt to model some features of growth at very low temperatures.
From the theoretical point of view there are two main phenomena which need to be simulated when trying to model crystal
growth: deposition and diffusion between subsequent depositions.
While the atom-surface impact can be modeled directly by MD, since the time-scale involved is of only a few ps, 
the real challenge is to describe the evolution of the system before the next impinging-atom landing. Here the difference
between MD accessible time scales and the experimental ones is huge: real deposition rates can be slower than 1ML/s, 
while MD ones are closer to $10^{7}$-$10^{9}$ ML/s (depending on the system size).
Only if the temperature is so low that virtually no activated events are possible between subsequent depositions, 
then the morphology of the  growing film is directly determined by the sequence of fast impacts, since atoms remain 
frozen to the final position reached after the impact energy is dissipated.
For the present system, the barrier for adatom diffusion along the fast, in-channel direction was calculated to 
be $\sim 0.28$ eV \cite{ferrandokmc}. Estimating the average residence time $\tau$ in a given site by the Arrhenius 
relation
\begin{equation}
\tau=\frac{1}{\nu_0} \exp{\left ( \frac{E}{k_B T} \right )}\; ,
\label{eq::arr}
\end{equation}
where $E$ is the activation energy of the mechanism bringing the system out of the site and $k_B$ is the 
Boltzmann constant, and by setting the frequency prefactor $\nu_0$ equal to the standard value $\nu_0 = 10^{13}$s$^{-1}$, 
we can easily estimate that for $T\lesssim 100$K, $\tau \gtrsim 10$s, a value for which surface diffusion can be 
declared frozen (unless extremely slow-deposition experiments are carried out). In this sense, our submonolayer-growth simulations
should be regarded as representative of real substrate temperatures lower than 100K.
 This is confirmed by the comparison with the KMC simulations \cite{videcoq02}, which show that all diffusive 
processes are practically frozen at 100 K and below.

However when we tried to use in the simulations a temperature comparable with this estimated threshold, the
ultra-fast deposition rate occasionally induced the formation of artifical metastable structures (a problem which 
did not come out in the single-adatom
deposition case), with the whole system trapped in ultra-shallow minima which
would suddenly disappear if enough time would be available to evolve. It was sufficient to raise the 
temperature to $T=300$K in order to eliminate this problem, without introducing any real diffusive event \cite{nota}.
Let us emphasize once again that a realistic modeling of growth at $T=300$K requires to leave the simple MD approach. Ideally, one should introduce the here presented
impact-driven mechanisms within a simulation scheme able to reach extended time scales. Kinetic Monte Carlo, or accelerated MD  techniques \cite{annurev} could be the right choice for a significant extension of our work.
In spite of the above illustrated limitations, we stress that all of the observations on the effects of deposition-parameters depending steering and transient mobility remain valid, since the whole set of simulations was run under the same high-rate conditions.

\section{Conclusions}
\label{sec::conclusions}

In this paper we have analyzed the effects of steering and transient mobility on deposition and
low-temperature submonolayer growth of Ag/Ag(110).

For deposition at normal incidence on a flat surface, the simulations have not revealed any significant 
transient mobility of the adatom, which stops after impact in the target cell in about 90\% of cases even 
when the initial energy is of $1$ eV. The incoming kinetic energy is dissipated in the substrate, without
giving rise to any directed motion in the surface plane. On the contrary, for grazing incidence along the 
in-channel direction, there is an interplay between steering and transient mobility, which is now possible
beacause the initial velocity has already a component on the surface plane.
Due to steering, the incoming atoms tends to fall shorter with respect to the target cell, 
while transient mobility has the opposite effect, pushing the atom to travel further by a few hopping
events that occur before thermalisation.

Steering and transient mobility have siginificant effects also on submonolayer growth. 
Normal incidence deposition causes some increase of the roughness of the film, because incoming atoms are 
attracted by already existing protrusion \cite{montalentiprb01}. However, this increased roughness effect
is much more important when depositing at grazing incidence with low initial kinetic energy.
 In this case, the incoming atom feels a much stronger steering effect, but without presenting significant
transient mobility after impact. This enhances the probability of forming dimers and small aggregates, on
 which a second layer can nucleate. For high-energy grazing incidence, the effects of transient mobility
 come into play.
Now, longer chains are formed on the surface at the expense of dimers. Moreover, roughness
 decreases because there is a smaller probability of nucleating a second layer, due to transient 
mobility causing interlayer downward processes.

Our results have thus shown the possibilty of tuning growth morphologies by changing the initial energy
 and the angle of incidence of incoming atoms already in the submonolayer regime. We expect that these effects would
 be even more important in multilayer growth, especially for grazing incidence deposition, due to the appeareance 
of shadowing effects besides steering and transient mobility. Work is in progress on this subject.



\end{document}